# Probing the wavefunctions of correlated states in magic angle graphene

Zhiming Zhang[1], Rachel Myers[1], Kenji Watanabe[2], Takashi Taniguchi[2], Brian J. LeRoy*[1]
[1]*Physics Department, University of Arizona, 1118 E 4th Street, Tucson, AZ 85721, USA.*
[2]*National Institute for Materials Science, Namiki 1-1, Tsukuba, Ibaraki 305-0044, Japan.*

Using scanning probe microscopy and spectroscopy, we explore the spatial symmetry of the electronic wavefunctions of twisted bilayer graphene at the "magic angle" of 1.1 degrees. This small twist angle leads to a long wavelength moiré unit cell on the order of 13 nm and the appearance of two flat bands. As the twist angle is decreased, correlation effects increase until they are maximized at the magic angle. At this angle, the wavefunctions at the charge neutrality point show reduced symmetry due to the emergence of a charge ordered state. As the system is doped, the symmetry of the wavefunctions change at each commensurate filling of the moiré unit cell pointing to the correlated nature of the spin and valley degeneracy broken states.

## I. INTRODUCTION

When two layers of graphene are twisted away from each other at the "magic angle", around 1.1 degree, the long wavelength moiré pattern leads to the folding of the band structure into a mini-Brillouin zone and the formation of ultra-flat bands [1–7]. As the twist angle approaches the magic angle, the bands become flatter and the ratio between the Coulomb interaction and their bandwidth increases leading to increased correlation effects [8–21] and broken symmetry states [22–24]. Superconductivity and correlated insulating states [25–29] have been discovered in magic angle twisted bilayer graphene (MATBG) when the flat bands are tuned to commensurate filling factors $\nu = 0, \pm1, \pm2, \pm3$ electron per moiré unit cell. For $\nu = \pm1(\pm 3)$, there is only one electron(hole) in one of the flat bands, hence both the spin and valley degeneracy must be broken while for even filling factors one of these degeneracies must be restored. Due to the correlated nature of these states, small changes in the filling factors can lead to dramatic changes in the wavefunctions. Transport studies [16,27–29] have revealed that the phenomenology of the commensurate states at odd filling levels are different from those at even filling levels. Theory calculations [16,22,30,31] suggest that odd and even filling levels will lead to different symmetry breaking phases.

Since there is still considerable theoretical uncertainty in the exact nature of the wavefunctions at each of the commensurate fillings, direct imaging of the wavefunctions can provide experimental insights for explaining the coherent states that have been observed in MATBG. For example, revealing the exact spatial distribution of flat band wavefunctions at different commensurate fillings can provide information such as anisotropy and localization properties. Our STM/STS study focuses on the difference in the wavefunctions between even and odd filling factors, which has not been explored by previous STM studies on MATBG [23,24,32–34]. We find that for the partially filled bands, the wavefunctions are localized on different part of the moiré unit cell. We further explore the properties of wavefunctions by quantitively extracting the anisotropy, localization, and radial distribution for various energy and filling levels, providing a comprehensive local characterization of flat bands in MATBG.

## II. EXPERIMENT

Figure 1(a) shows a schematic of the experimental setup. All measurements were performed in ultra-high vacuum at a temperature of 4.6 K. Samples were fabricated by a dry transfer technique with controlled rotational alignment between the two layers of graphene [35]. A bias voltage ($V_b$) applied between the tip and the sample is used for probing different energy levels, while the gate voltage ($V_g$) applied between the Si substrate and the sample is used for tuning the doping level of the MATBG. Figure 1(b) shows an STM topography image of a 1.10° MATBG moiré pattern, the twist angle is determined by measuring the moiré wavelengths in 3 different directions (L1, L2, L3) and using a uniaxial heterostrain model [24]. The bright spots in the topography image correspond to the AA sites of the moiré superlattice, other high symmetry sites are labeled as AB, BA. The atomic arrangements of these high symmetry points are illustrated in the schematic for a 5° moiré pattern in Fig. 1(c). Figure 1(d) shows the scanning tunneling spectroscopy (STS) on the AA and AB sites when the flat bands are fully occupied. The two sharp peaks between -0.05 V and 0 V in the spectroscopy of the AA site correspond to the strong local density of states (LDOS) of the flat bands, while the LDOS in this energy range is much weaker on the AB site, indicating that the flat bands are localized on the AA site [3,5].

## III. TUNNELING SPECTROSCOPY

To study the electronic properties of the flat bands at different doping levels, we performed STS measurements on the AA site as a function of both $V_b$ and $V_g$, as shown in Fig. 2(a) for a 1.07° twist angle moiré. For each $V_g$, the tip height was stabilized at a bias voltage of 0.1 V and tunnel current of 50 pA. Then the feedback circuit was switched off, a small ac voltage (1 mV) was applied to the bias voltage and the differential conductance dI/dV was measured as a function of bias voltage using lock-in detection. The tip gating effect [33] was corrected from the data, see Appendix 1 for details. When the two flat bands are fully filled ($V_g$ >14 V) or fully depleted ($V_g$ <-50 V), they are close together with a peak to peak separation of ~14 mV and their energy shifts linearly with respect to $V_g$ with a slope of ~3 mV per volt in $V_g$. In contrast, in the region where the flat bands are overlapping with the Fermi level (-50 V< $V_g$<14 V), the energy of the flat bands changes much slower with respect to $V_g$. On top of this, various new features of the flat bands emerge:

(1) The total band width of the flat bands, defined as the full width at the half maximum on either side of the flat bands, indicated in Fig. 2(b) and Fig. 2(c), is broadened in this region, shown by the black curve in Fig. 2(f). The overall broadening indicates stronger electron-electron interactions and the breakdown of a single particle picture [30,32]. Such broadening is not present in the spectroscopy of non-magic angle devices, as shown in Appendix 2.

(2) The separation between the two flat bands is enhanced near the charge neutrality point (CNP, defined as where the conduction band and valence band are equidistant from the Fermi level and have the same peak LDOS) around $V_g$ = -26 V, which is a result of stronger interlayer exchange interactions [30,32,33]. To illustrate that this effect is strongest near the magic angle of 1.1°, we measured the total band width at CNP ($W_{CNP}$) and when both bands are fully depleted ($W_{FD}$) from dI/dV measurements of different twist angle devices. The total band width as a function of angle is plotted in Fig. 2(d), while the band width decreases with the twist angle, the difference between $W_{CNP}$ and $W_{FD}$ increases at small angle. As shown in Fig. 2(e), the ratio between $W_{CNP}$ and $W_{FD}$ shows a peak around 1.07°, confirming that the exchange interaction is indeed strongest around the magic angle. Other angle dependent quantities such as separation between the two flat bands, individual band widths and the separation between the flat bands and remote bands are presented in Appendix 3.

(3) Near the CNP, the width of the upper conduction band is wider than the lower valence band in the p-doped region ($V_g$ < -26 V) and vice versa in the n-doped region ($V_g$ > -26 V), see the red and blue curves in Fig. 2(f). The spectral weight redistribution between the two flat bands as a function of doping is consistent with previous experiments suggesting a correlated charge ordered phase [23], Mott-insulating states as well as superconductivity near CNP have also been observed in transport measurements [28].

(4) In the range of -15 V< $V_g$<13 V when the lower band is fully filled and the upper band is partially filled, both bands show various distortions. Dips appear in the LDOS in the upper band at certain doping levels and the lower band is broadened when the upper band is partially filled. To better identify the location of the dips and compare our STS results with transport experiments [25–29], we plot the LDOS at the Fermi level as a function of $V_g$ in Fig. 3(a), which is a horizontal line cut from Fig. 2(a) at zero bias voltage. Consistent with transport measurements [25–29], strong insulating states appear when the bands are fully filled ($V_g$ >14 V), fully depleted ($V_g$ <-50 V) and around the CNP (-46 V<$V_g$ < -20 V). Additional dips in the LDOS show up around $V_g$ = -12 V, -1 V and 9 V, similar to the insulating states with filling factors of $\nu$ = 1, 2 and 3 observed in transport measurements [25–29]. From the individual dI/dV curves at these gate voltages in Fig. 3(b), a soft gap like structure appeared at the Fermi level for $V_g$ = -12V ($\nu$ = 1) and $V_g$ = -1V ($\nu$ = 2) but not for $V_g$ = 9V ($\nu$ = 3).

Fig. 3(c) illustrates the evolution of the two flat bands as a function of filling levels, pink and gray rectangles correspond to the upper and lower bands respectively, solid and empty dots correspond to electrons and holes that are occupying the flat bands. At $\nu$ = 0, both bands are broadened, and their separation is enhanced comparing to $\nu$ = ±4. When the upper band is partially filled ($\nu$ = 1, 2, 3), the upper band is split into two bands on either side of the Fermi level, meanwhile the lower band is further broadened.

The asymmetry of the dI/dV as a function of $V_g$ as well the missing a gap at ν = 3 is because of tip-induced band bending [32]. With different tip conditions, we observed symmetric gate dependent spectroscopy with no tip-induced bending as well as asymmetric gate dependent spectroscopy with tip-induced bending in a different direction, as shown in Appendix 4. The dI/dV signal in Fig. 2(a) is much stronger when the flat bands are fully filled, this is because we are setting the tip at a positive voltage (0.1V) when starting to take the spectroscopy. When the bands are fully filled, there is almost no density of states present from the Fermi level up to the positive tip voltage, thus the tip height will be stabilized closer to the sample in order to reach the set current, which in turn increases the dI/dV signal. By setting the starting voltage at a negative voltage, the dI/dV signal is stronger when the flat bands are fully depleted, see Appendix 4.

## IV. DENSITY OF STATES MAPS

Although gap opening and insulating behavior has been observed at all commensurate fillings: ν = 0 (CNP), ±1, ±2 and ±3. The intrinsic gap driving mechanism for these states can be very different [13,16,27,30,31]. Parallel field dependent transport measurements [27] have found that the ν = ±1 and ν = ±3 states are spin-polarized while the ν = ±2 states are not. A ferromagnetic state [29] and intrinsic quantized anomalous Hall effect [20] have been found in the ν = 3 state, Chern insulating states have been discovered for the ν = ±1 states [28]. To highlight the difference between the different commensurate filling states, which have not been addressed in the previous STM studies on MATBG [23,24,32,33], we measure the spatial profile of the wave functions at various energy and doping levels by mapping the LDOS with different $V_b$ and $V_g$.

The LDOS maps at the energies of the flat bands for different commensurate fillings are shown in Fig. 4, each one of them are averaged over several unit cells and plotted in the Wigner-Seitz cell of the moiré lattice centered at the AA site. The dI/dV value is normalized so that the summation of dI/dV over the unit cell is one, then the average value is subtracted, thus the average value will always appear as white color in the color plot, independent of the setting of the color scale. In order to compare these LDOS maps between different fillings, all images are plotted under the same color scale. The strain directions that are extracted from the uniaxial heterostrain model [24] are plotted in Fig. 4(a) with respect to the unit cell, where the strain percentage was found to be ε = 0.17% using the Poisson ratio δ = 0.16 for graphene. Figure 4(b) shows the unit cell topography of the same area, which has almost the full $C_6$ symmetry, only slightly altered by the strain. In contrast, the LDOS maps at the CNP, figure 4(c), show reduced symmetry with the symmetry axes perpendicular to each other between the lower band ($V_b$ = -18 mV) and the upper band ($V_b$ = 14 mV), which is a result of a charge ordered state [23]. When both bands are fully filled, figure 4(d), similar broken symmetry still exists but the shape of the wavefunctions for the lower ($V_b$ = -106 mV) and upper ($V_b$ = -90 mV) bands are not as perpendicular to each other as they are in Fig. 4(b), indicating the absence of a charge ordered state [23]. LDOS maps at other energies and doping levels, as well as a qualitative study of the anisotropy and localization are present in Appendix 5. Interestingly, the nature of the wavefunctions changes dramatically when the bands are partially filled, as shown in Fig. 4(e) and Fig. 4(f) for ν = 1 and ν = 2. The wavefunctions of the fully filled lower band ($V_b$ = -32 mV for ν = 1, $V_b$ = -23 mV for ν = 2) are much less localized compared to Fig. 4(b) and Fig. 4(e). The width of the bands in our spectroscopy measurements is related to their flatness in momentum space. The larger the range of momenta for a given state, the more localized it will be in real space. Thus, the delocalization when the upper band is partially filled is related to the broadening of the lower band as shown previously in Fig. 2. The delocalization of the wavefunction further supports that the broadening of the flat bands is intrinsic, not from tip induced effects.

For the partially filled upper band, the wavefunctions are no longer localized at AA sites when ν = 1, the occupied state ($V_b$ = -7 mV) and the unoccupied state ($V_b$ = 5 mV) are localized on opposite locations somewhere between AA (center of the unit cell) and the boundary of the unit cell. On the other hand, the wavefunctions of the upper band for ν = 2 are still localized on the AA sites. Compared to Fig. 4(b), these wavefunctions do not have rotational symmetry and further they are not perpendicular to each other, which provides further evidence that the perpendicularity of wavefunctions at the CNP originates from a charge ordered state [23], since both the occupied and unoccupied wavefunctions here at ν = 2 are from the upper band. The stark difference of the conduction band wave functions between ν = 1 and ν = 2 fillings indicates that they are intrinsically different correlated states, which could be attributed to the fact that the ν = 1 state is both spin and valley polarized while the ν = 2 state can only either be spin-polarized or valley polarized [16,27,28,30,31]. Theoretical studies focusing on the spatially resolved calculation of wavefunctions with different degeneracies could provide understanding of the broken symmetries in this system.

## V. CONCLUSIONS

In summary, we have shown that the wavefunctions of the insulating state at CNP is consistent with stripe charge order [23]. The insulating states at $\nu = 1$ and $\nu = 2$ have distinct wavefunctions, their lower valence band wavefunctions are delocalized while their conduction band wavefunctions are localized on different sites. The coherent-driven broken symmetry states that we have observed are consistent with transport measurements [25–29] and previous STM studies [23,24,32,33] but highlight the intrinsic difference between the correlated insulating states at different commensurate fillings, which was not discovered previously. By direct measurement of the LDOS maps and the localization of these wavefunctions, we visualized the spatial dependence of the flat band wavefunctions as a function of doping level, providing a deeper understanding towards the nature of correlated states in MATBG at various commensurate fillings.

## ACKNOWLEDGMENTS

The authors thank Allan H. MacDonald for valuable theoretical discussions. The work at the University of Arizona was supported by the National Science Foundation under grants DMR-1708406 and EECS-1607911 and the Army Research Office under Grant No. W911NF-14-1-0653. K.W. and T.T. acknowledge support from the EMEXT Element Strategy Initiative to Form Core Research Center, Grant Number JPMXP0112101001 and the CREST(JPMJCR15F3), JST.

## APPENDIX

### 1. Tip gating correction for dI/dV measurements

In the moiré region close to the magic angle, quantum dots can be induced by the formation of insulating states [33], they appear as constant density resonance lines in Fig. 5, indicated by the red arrows. In the uncorrected image, Fig. 5(a), the constant density lines are slopped due to tip gating effects. In Fig. 5(b) we correct the tip gating effect by shifting the dI/dV curves to make the constant density lines vertical, thus the bias voltage and carrier density are decoupled in the corrected image. The tip gating correction does not change the appearance of the correlated effects as we discussd in the main text, however, it enables more accuate measurement of band widths.

### 2. Non-magic angle spectroscopy

To show that the correlation effects we discussed in the main text are only present in the moiré flat bands close to the magic angle, Fig. 6 shows the gate dependent STS spectroscopies for 1.30° and 2.31° twist angles. Consistent with pervious observations [36], the van Hove singularity (VHS) peaks are still present for both areas, however, in contrast to the magic angle area, there is no broadening of the peaks when they are partially filled. The separation between the peaks is almost constant for all gate voltages, and no significant distortion of the peaks occurs near commensurate fillings.

### 3. Angle dependence of band widths and separation between bands

In the main text, we have shown the total band width of both flat bands decreases with angle and the difference between the total band width at the CNP and FD is maximum around the magic angle. Figure 7(a) and (b) shows that the band widths of the individual bands also decrease with angle, but the difference between the CNP and FD do not show clear indication of enhancement around the magic angle, suggesting that the enhanced separation at CNP originates from inter-band interactions. Figure 7(c) shows the VHS separation decreases with angle, the ratio of VHS separation at the CNP compared to FD also increases at small angle as shown in Fig. 7(d). Figures 7(e) and (f) show the gap between the valence flat band and lower dispersive band and the gap between the conduction flat band and the higher dispersive both decrease with angle, consistent with theoretical predictions [6] and previous STS measurements [33].

### 4. Tip-induced asymmetry in gate dependent dI/dV measurements

The gate dependent STS measurements often show asymmetric shapes in dI/dV as a function of gate voltages, this tip band bending effect can be explained by the difference in the work function between the STM tip and the sample [32]. With different tip conditions, we observed symmetric spectroscopy in Fig. 8(a) and asymmetric spectroscopy bending towards the opposite direction in Fig. 8(b) as compared with Fig. 5(b).

The strength of dI/dV is also asymmetric as a function of gate voltage, this is affected by the setpoint of the bias voltage as discussed in the main text. Figure 9(a) and Figure 9(b) shows the gate dependent STS spectroscopy on the same spot on the sample and with the same tip conditions but with setpoints of opposite sign, as a result, the dI/dV is enhanced in different gate voltage ranges of the image.

## 5. LDOS maps at different energy and filling levels

In addition to Fig. 4, Figure 10(a) shows a series of LDOS maps around the flat band energies for ν = 1. To quantify the $C_3$ rotational symmetry breaking, we follow the definition of energy-dependent anisotropy $A(E)$ as [24]:

$$A(E) = \frac{1}{3}\sum_{unit\ cell}\left(\frac{|I_{0^\circ}(E) - I_{120^\circ}(E)|}{I_{0^\circ}(E) + I_{120^\circ}(E)} + \frac{|I_{120^\circ}(E) - I_{240^\circ}(E)|}{I_{120^\circ}(E) + I_{240^\circ}(E)} + \frac{|I_{240^\circ}(E) - I_{0^\circ}(E)|}{I_{240^\circ}(E) + I_{0^\circ}(E)}\right)$$

where $I_{0^\circ}(E)$ is the spatial LDOS profile when the image is aligned with the unit cell, $I_{120^\circ}(E)$ and $I_{240^\circ}(E)$ correspond to the spatial LDOS profile when $I_{0^\circ}(E)$ is rotated 120° and 240° respectively. As seen in Fig. 10(b), when the wavefunctions are delocalized near the filled valence flat band energies ($V_b$ = -36 ~ -32 mV), their anisotropy is comparable to the anisotropy of the topography image (Fig. 2(b)), which is 2.94%. In contrast, the anisotropies for the partially filled conduction flat band wavefunctions are much larger. To quantify how centered the wavefunctions are at the AA site, we define the radial distribution by averaging the LDOS over a circle at each radial distance from the center of the AA site, then normalized so that the summation over the full unit cell is 1. Figure 10(c) shows the radial distribution of the wavefunctions, gray area means that the LDOS maps are not available in Fig. 10(a). Here the delocalized wavefunctions show uniform distribution, while the unoccupied valence flat band states ($V_b$ = 1 ~ 5mV) are gradually shifted away from the AA sites. To quantify the overall localization of the wavefunctions, we plot in Fig. 10(d) a histogram at each bias voltage of the normalized LDOS for each wavefunction in Fig. 10(a), together with the STS spectroscopy taken at the center of the AA site plotted against the right axis. When the histogram is more concentrated as a function of LDOS, the wavefunction is more uniform or delocalized, because most pixels in the image have the same intensity. When the histogram has a broader distribution of LDOS values, there is more contrast in the image, which means more contrast or more localization. Figure 10(d) shows that the filled valence flat band states are strongly delocalized, while the partially filled conduction flat band states are still localized.

The same analysis is applied for ν = 0, 2, 3, 4 and -4 and shown in Figs. 11 – 15. For ν = 0, 4 and -4, the wavefunctions are always localized on the AA sites. For ν = 2 and 3, the filled valence flat band is delocalized, consistent with our findings in the main text. The partially filled conduction flat band is still localized on AA for ν = 2 but localized on a different location for ν = 3. Due to tip-induced band bending, all flat band energies are below the Fermi level for ν = 3. However, their wavefunctions still show similar behavior as the wavefunctions for ν = 1.

*leroy@arizona.edu

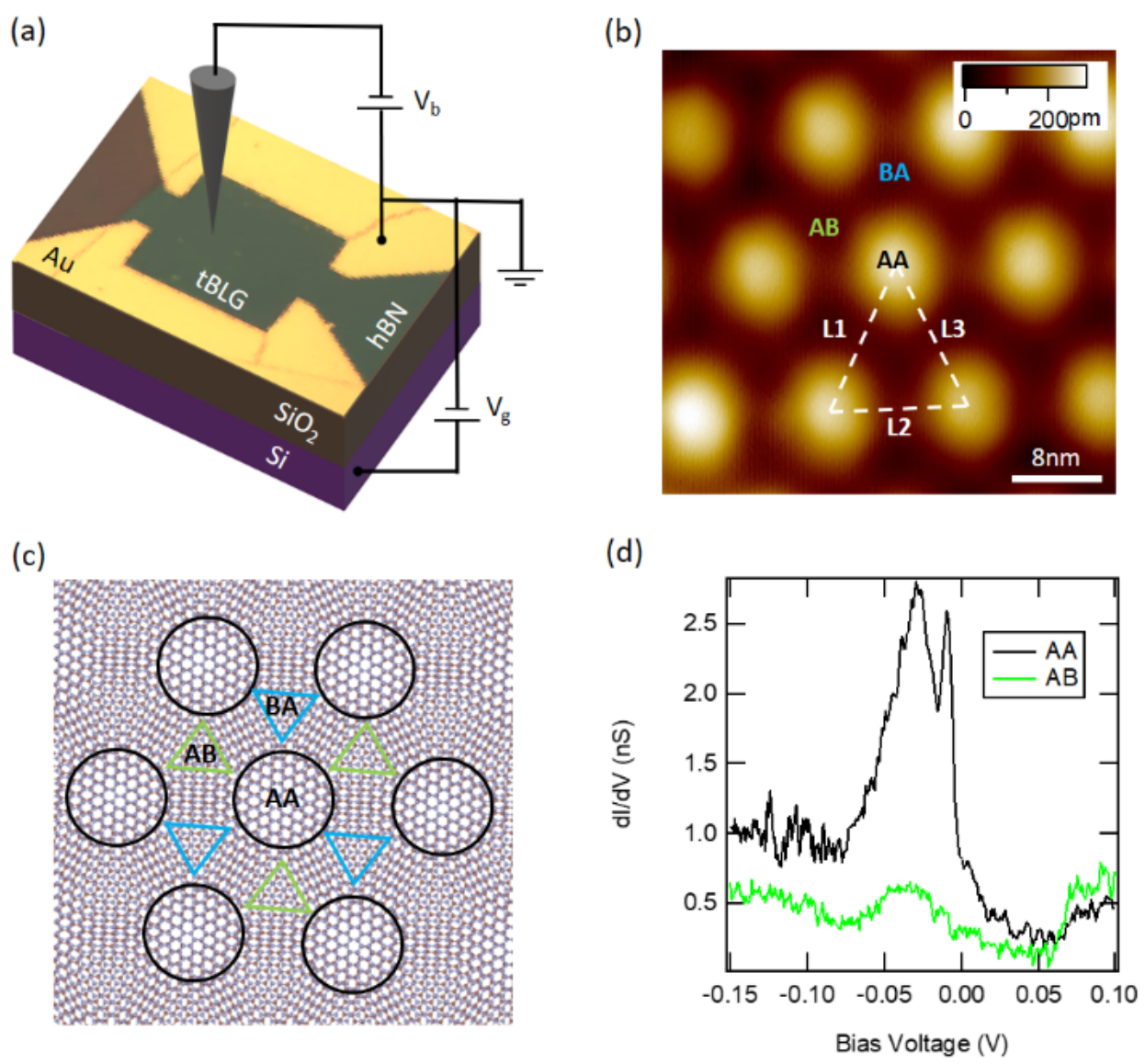

FIG. 1. (a) Schematic of the experimental setup showing the STM tip and an optical microscope image of the measured MATBG sample. (b) Atomic resolution STM topography of the 1.10° moiré superlattice. (c) Schematic of different atomic stacking arrangements arising from the twist of the two layers of graphene. (d) dI/dV spectroscopy of the MATBG on AA and AB sites of 1.10° moiré when the flat bands are fully filled, where $V_g$ = 8 V.

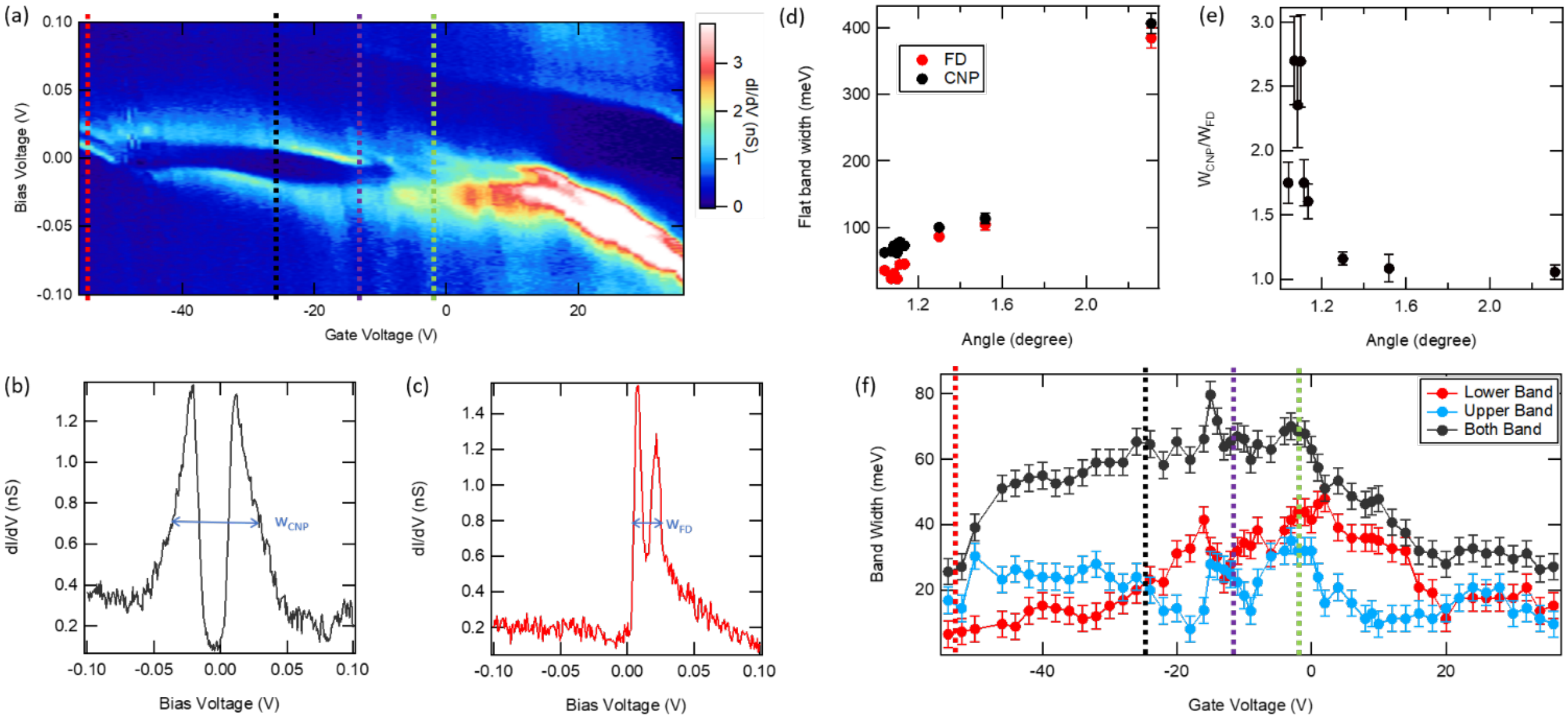

FIG. 2. (a) Gate dependent STS spectroscopy on the AA site of the 1.07° moiré superlattice. Dashed lines indicate the position of commensurate fillings: ν = -4 (red), ν = 0 (black), ν = 1 (purple), ν = 2 (green). (b) dI/dV curve at CNP, $V_g$ = -26 V. (c) dI/dV curve at FD, $V_g$ = -54 V. (d) Flat band width at two different fillings as a function of twist angle. (e) The ratio of flat band width at two different fillings as a function of twist angle. (f) The flat band width extracted from (a) as a function of $V_g$.

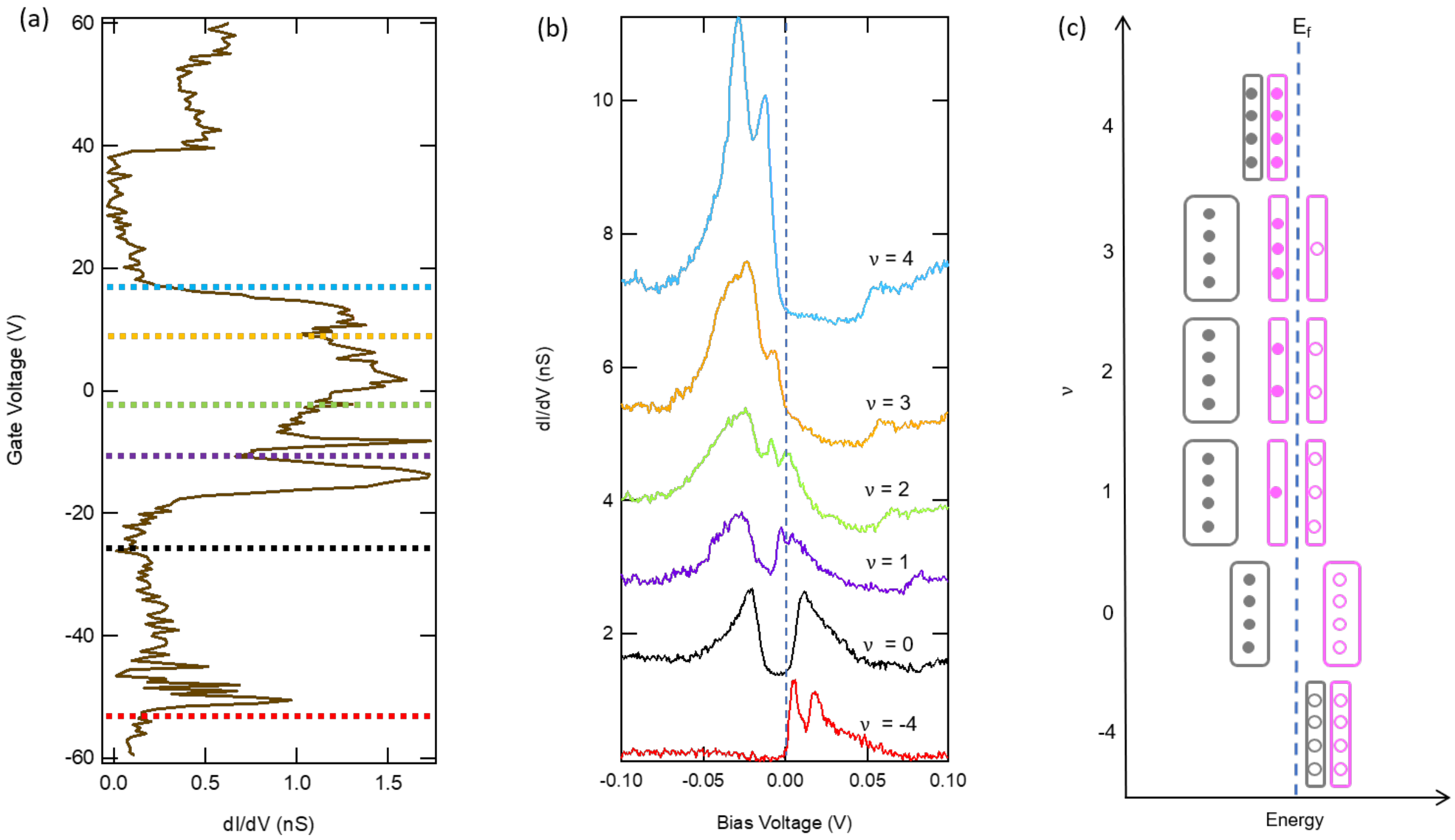

FIG. 3. (a) Gate dependent LDOS at the Fermi level on the 1.07° moiré superlattice, colored dashed line corresponding to different commensurate fillings. (b) dI/dV curves at different commensurate fillings, vertical dashed line marks the fermi level. (c) Illustration of the evolution of the flat bands at different filling levels. Pink and gray correspond to the upper band and lower band respectively while solid dots are electrons and circles are holes.

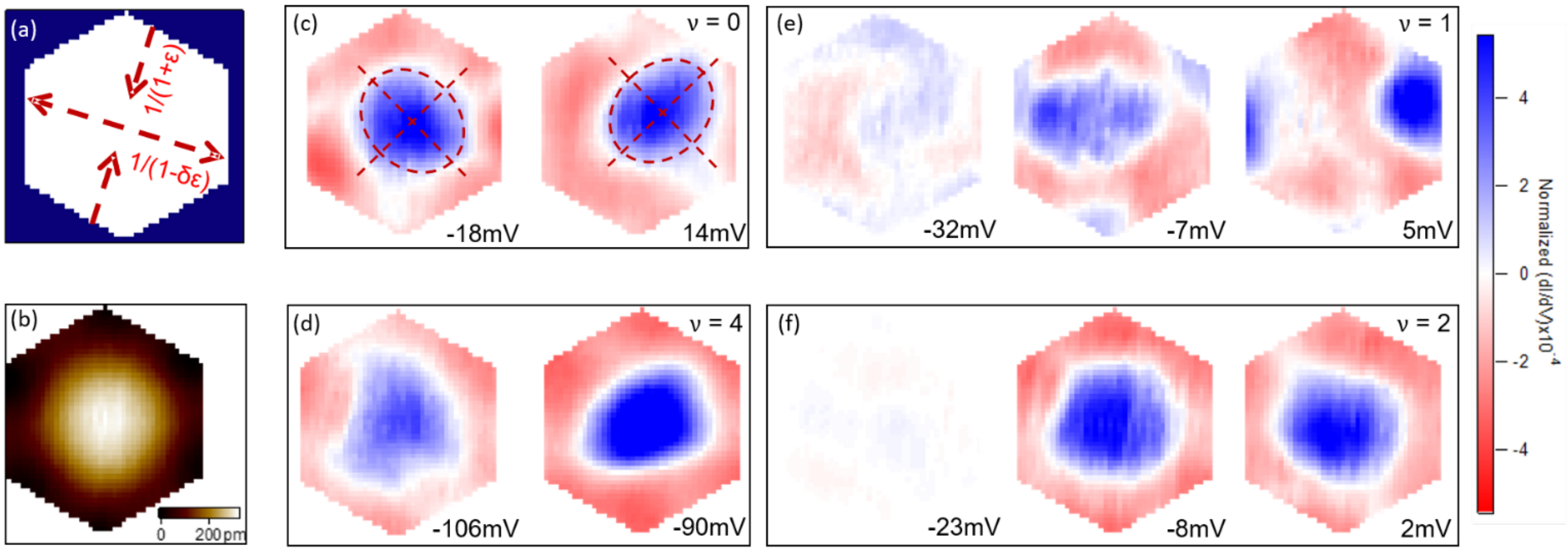

FIG. 4. (a) Uniaxial strain directions relative to the unit cell. (b) Topography of the unit cell. (c) LDOS maps at CNP ($\nu = 0$), dashed ovals highlight the shape of the wave functions and dashed lines shows the symmetry axes. (d) LDOS maps when the upper band is filled with fully filled ($\nu = 4$). (e) LDOS maps when the upper band has one electron ($\nu = 1$). (f) LDOS maps when the upper flat band is half filled ($\nu = 2$).

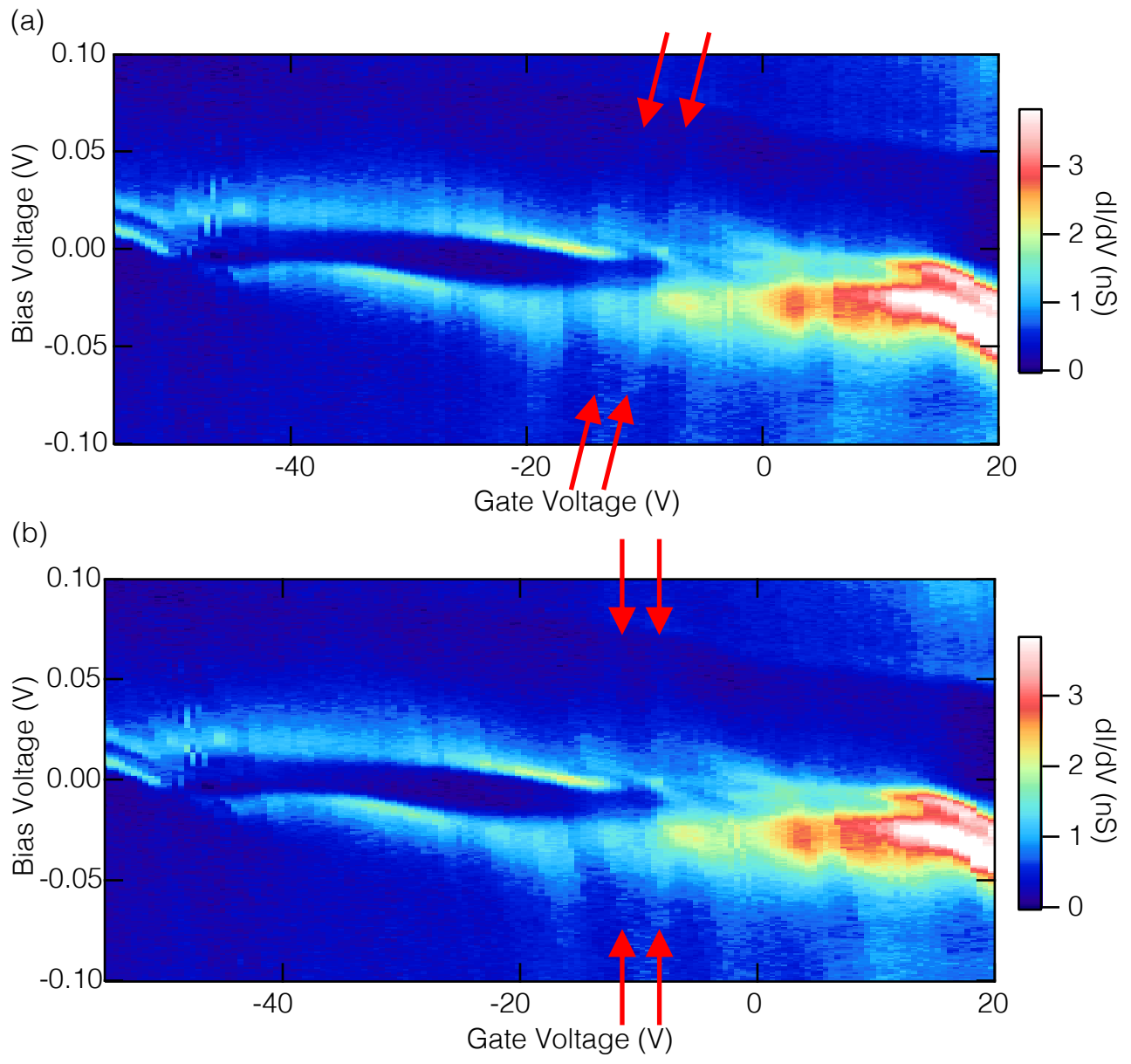

FIG. 5. (a) Gate dependent STS spectroscopy on the AA site of the 1.07° moiré superlattice before the tip gating correction. (b) Gate dependent STS spectroscopy after the bias voltage is decoupled from the carrier density.

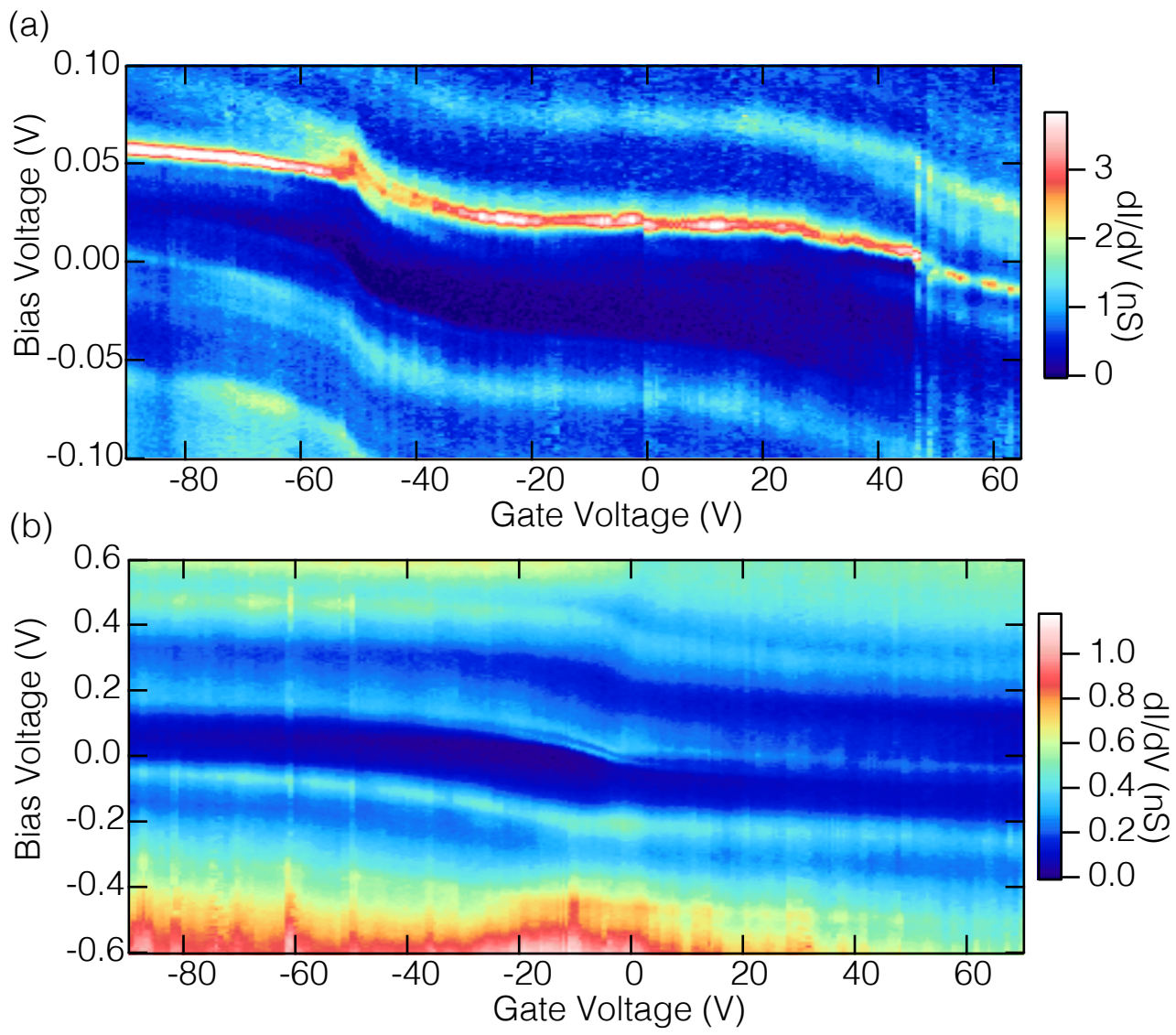

FIG. 6. (a) Gate dependent STS spectroscopy on the AA site of the 1.30° twist angle device. (b) Gate dependent STS spectroscopy on the AA site of the 2.31° twist angle device.

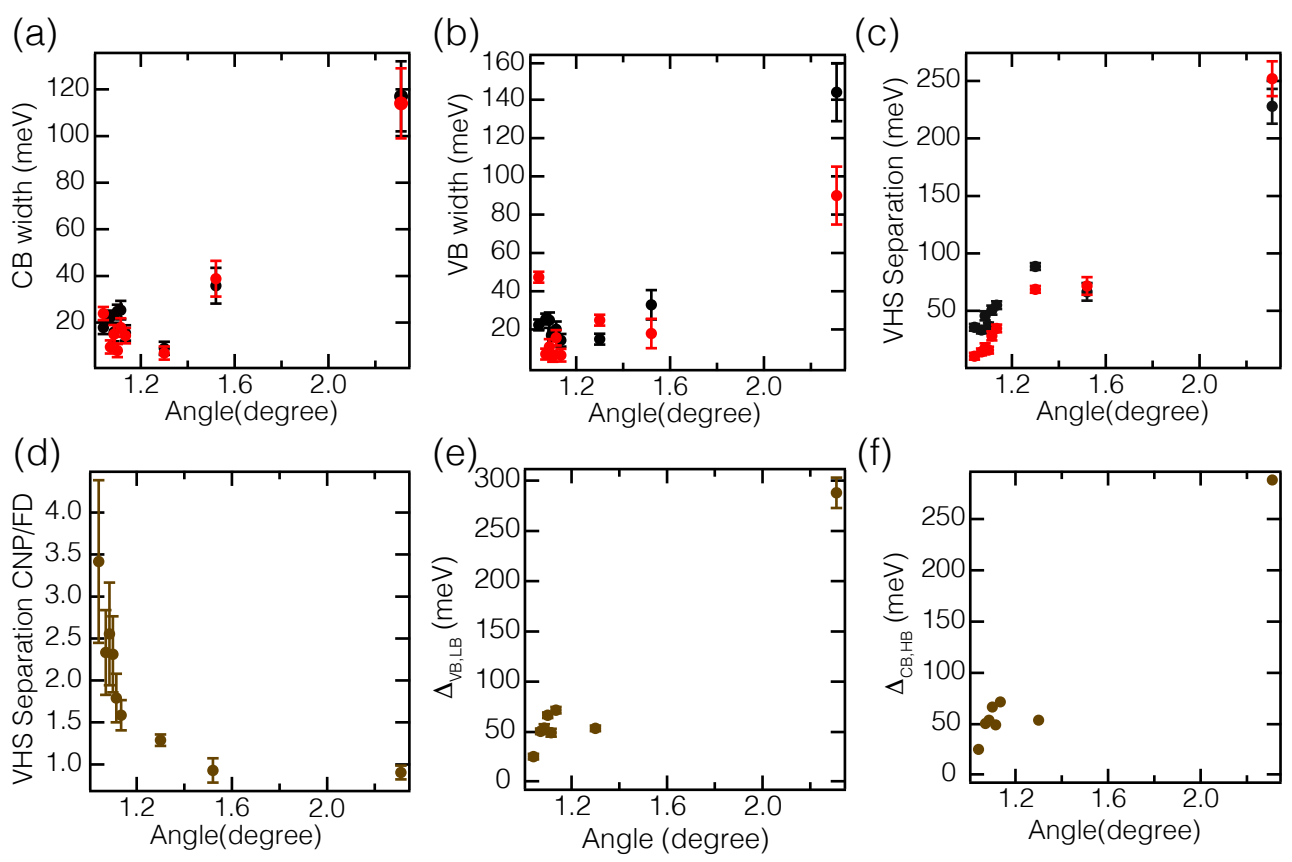

FIG. 7. Angle dependence of (a) the conduction band width, (b) the valence band width, and (c) the VHS separation. The red circles are when the band is fully depleted, and the black circles are at the CNP. Angle dependence of (d) the ratio of the VHS separation between CNP and FD, (e) the gap between the valence flat band and the lower dispersive band, and (f) the gap between the conduction flat band and the higher dispersive band.

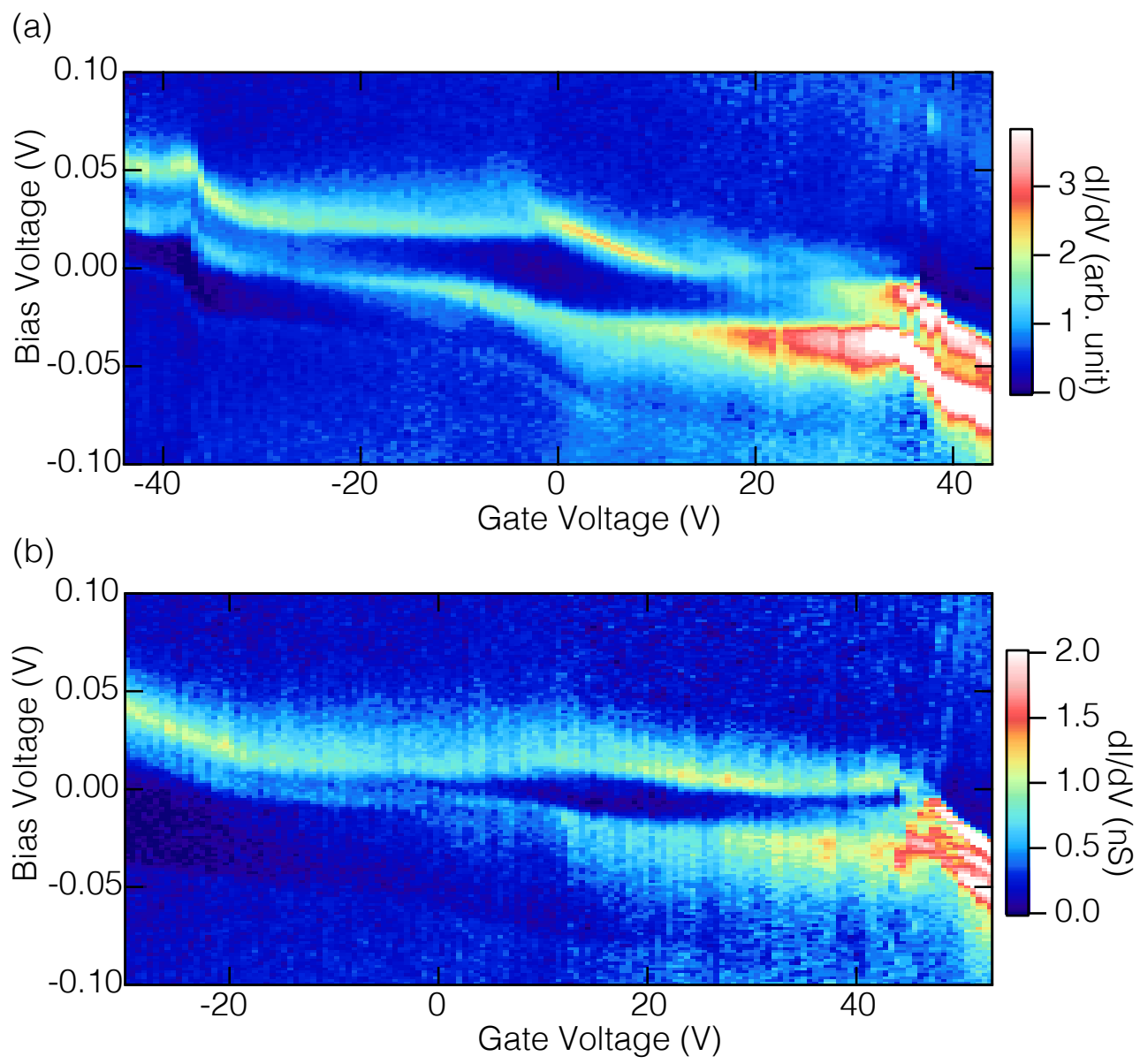

FIG. 8 (a) Gate dependent STS spectroscopy on the AA site of the 1.11° moiré superlattice without strong tip band bending effect. (b) Gate dependent STS spectroscopy on the AA site of the 1.09° moiré superlattice with strong tip band bending effect.

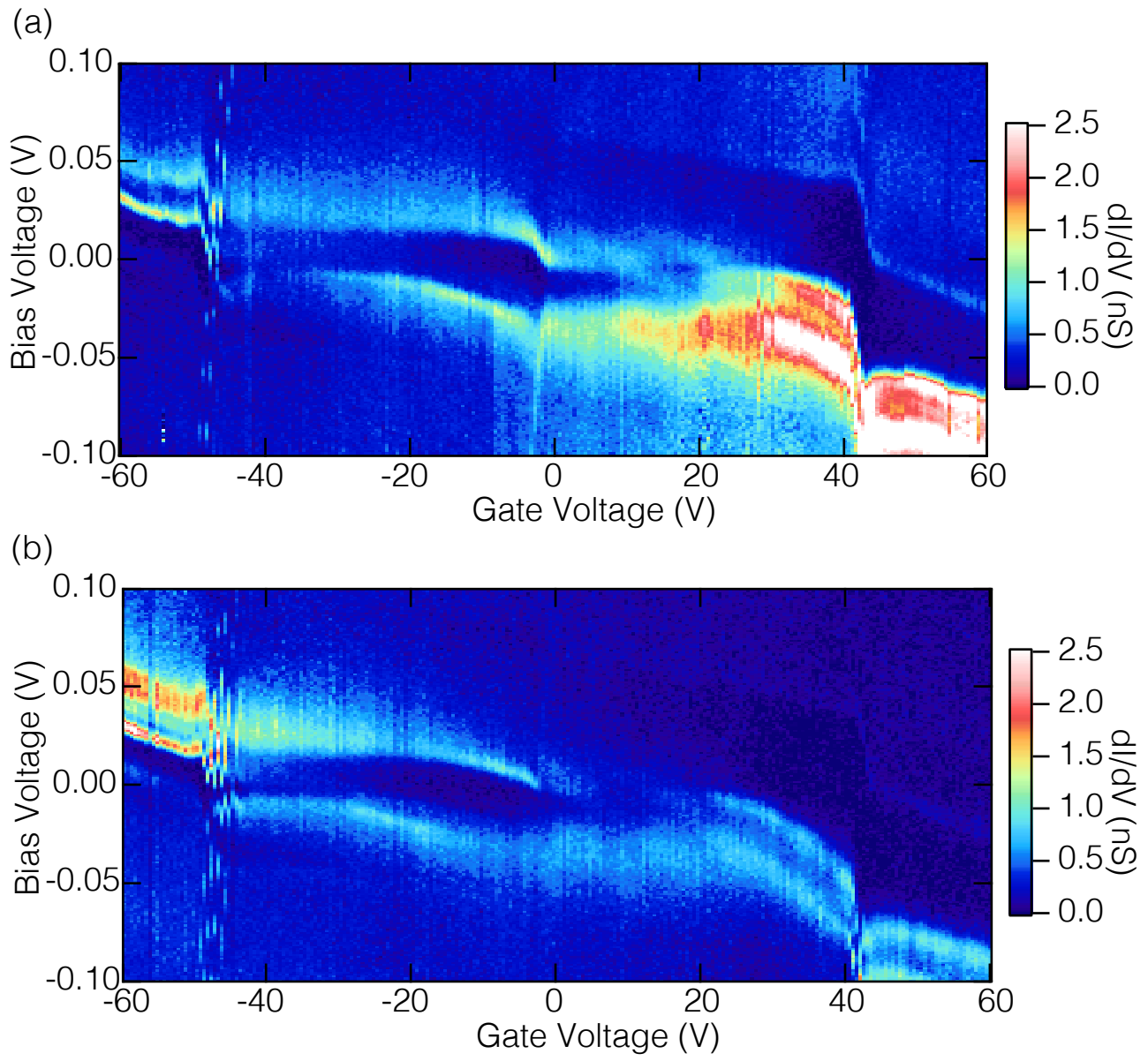

FIG. 9. (a) Gate dependent STS spectroscopy on the AA site of the 1. 08° moiré superlattice with a setpoint voltage of 0.15 V. (b) Gate dependent STS spectroscopy on the AA site of the 1. 08° moiré superlattice with a setpoint voltage of -0.15 V.

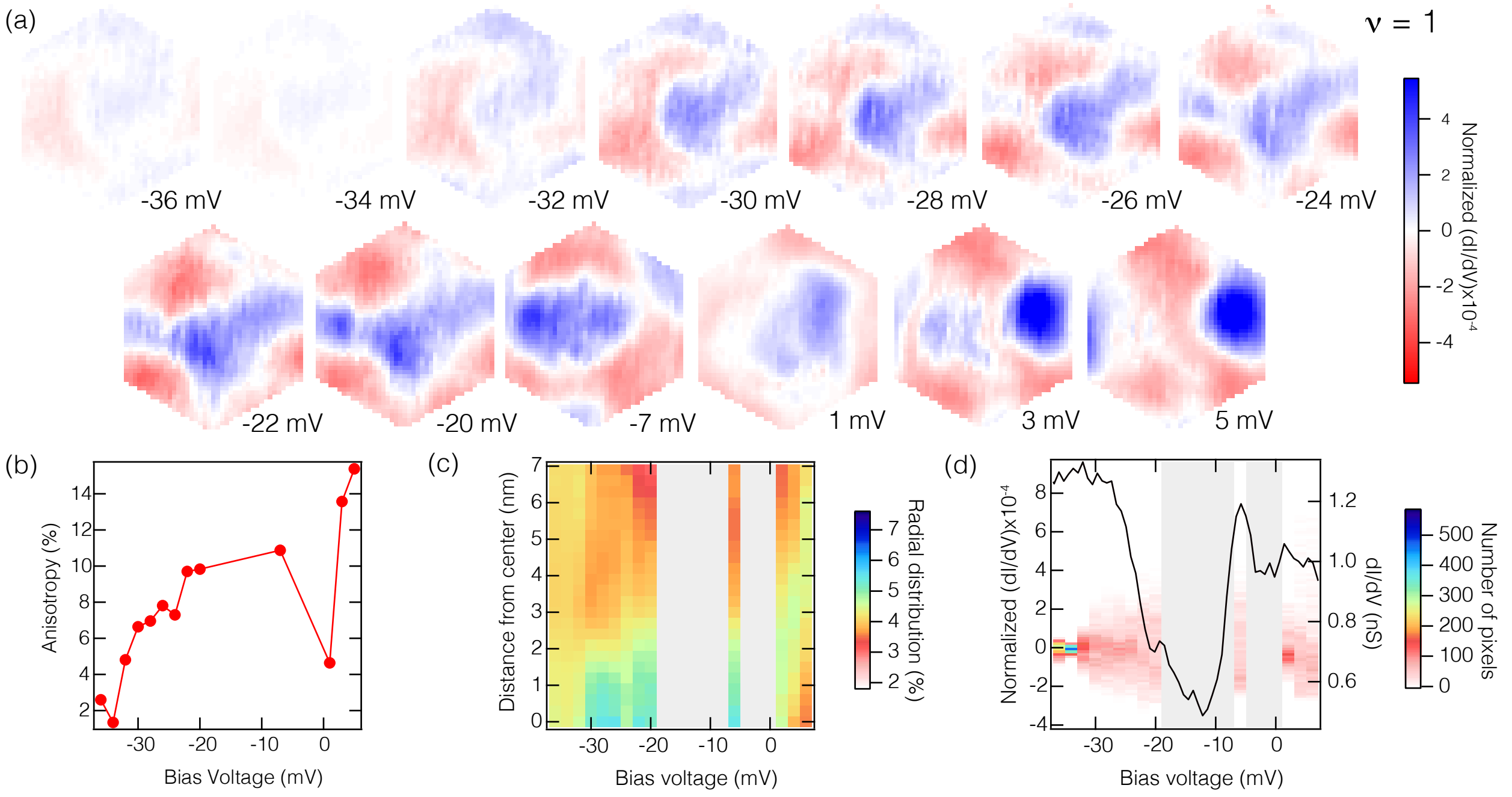

FIG. 10. (a) LDOS maps at different energies for ν = 1. (b)Anisotropy extracted from (a) as a function of bias voltage. (c) Radial distribution extracted from (a) as a function of bias votlage. (d) Color plot plotted against left axis: histogram of (a), black curve ploted against right axis: STS spectroscopy on the center of AA site.

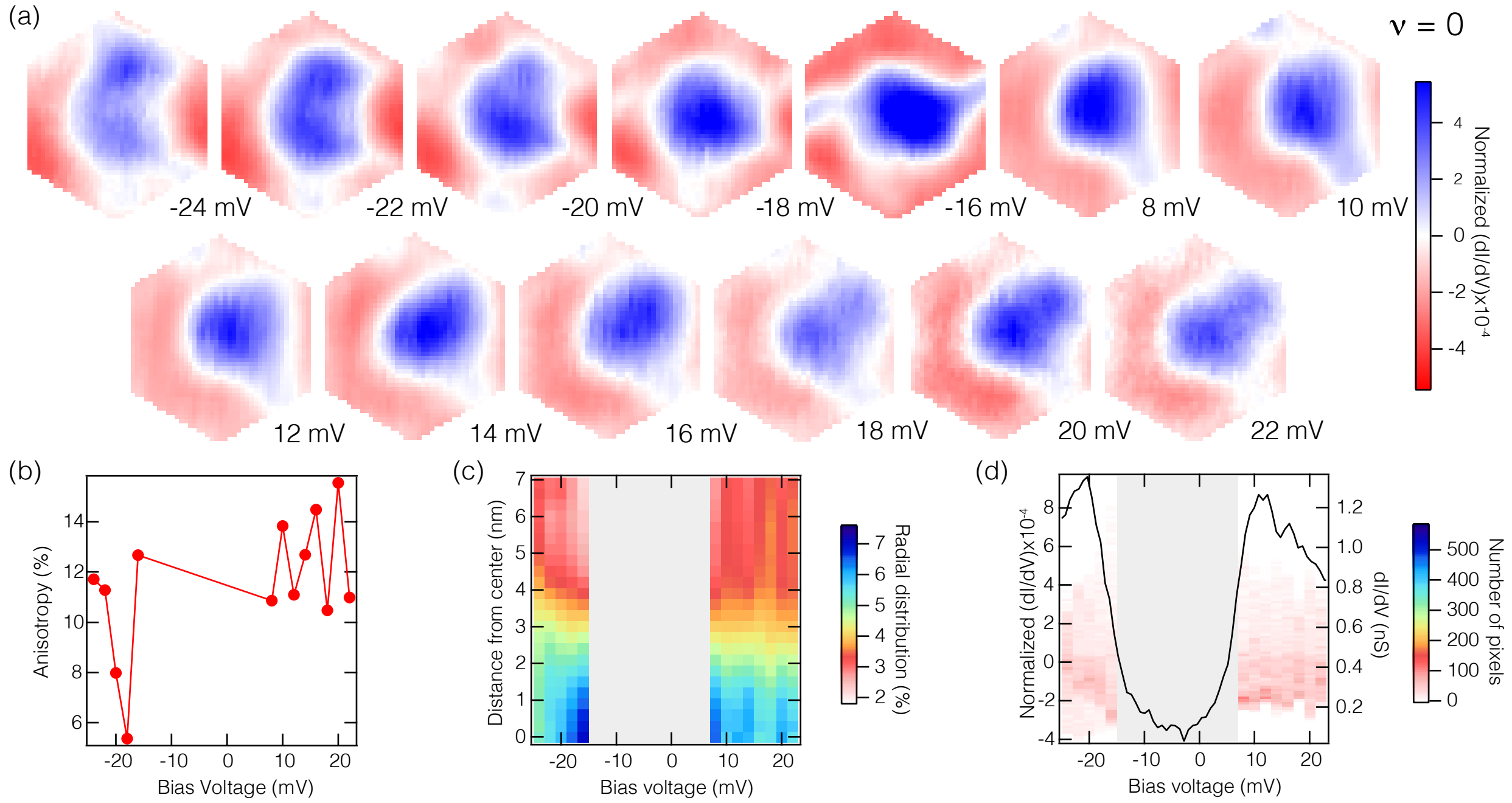

FIG. 11. (a) LDOS maps at different energies for ν = 0. (b)Anisotropy extracted from (a) as a function of bias voltage. (c) Radial distribution extracted from (a) as a function of bias voltage. (d) Color plot plotted against left axis: histogram of (a), black curve plotted against right axis: STS spectroscopy on the center of AA site.

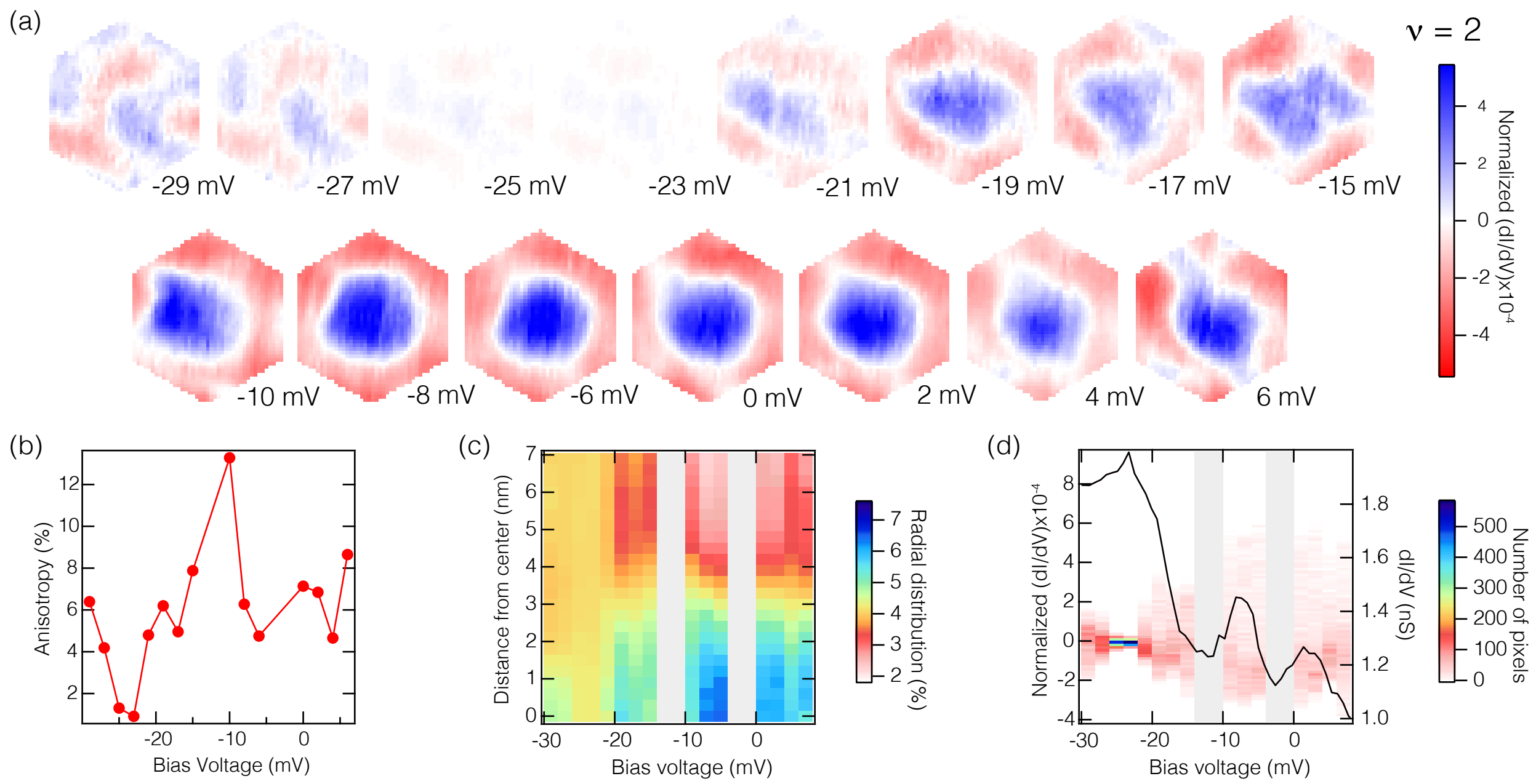

FIG. 12. (a) LDOS maps at different energies for ν = 2. (b)Anisotropy extracted from (a) as a function of bias voltage. (c) Radial distribution extracted from (a) as a function of bias voltage. (d) Color plot plotted against left axis: histogram of (a), black curve plotted against right axis: STS spectroscopy on the center of AA site.

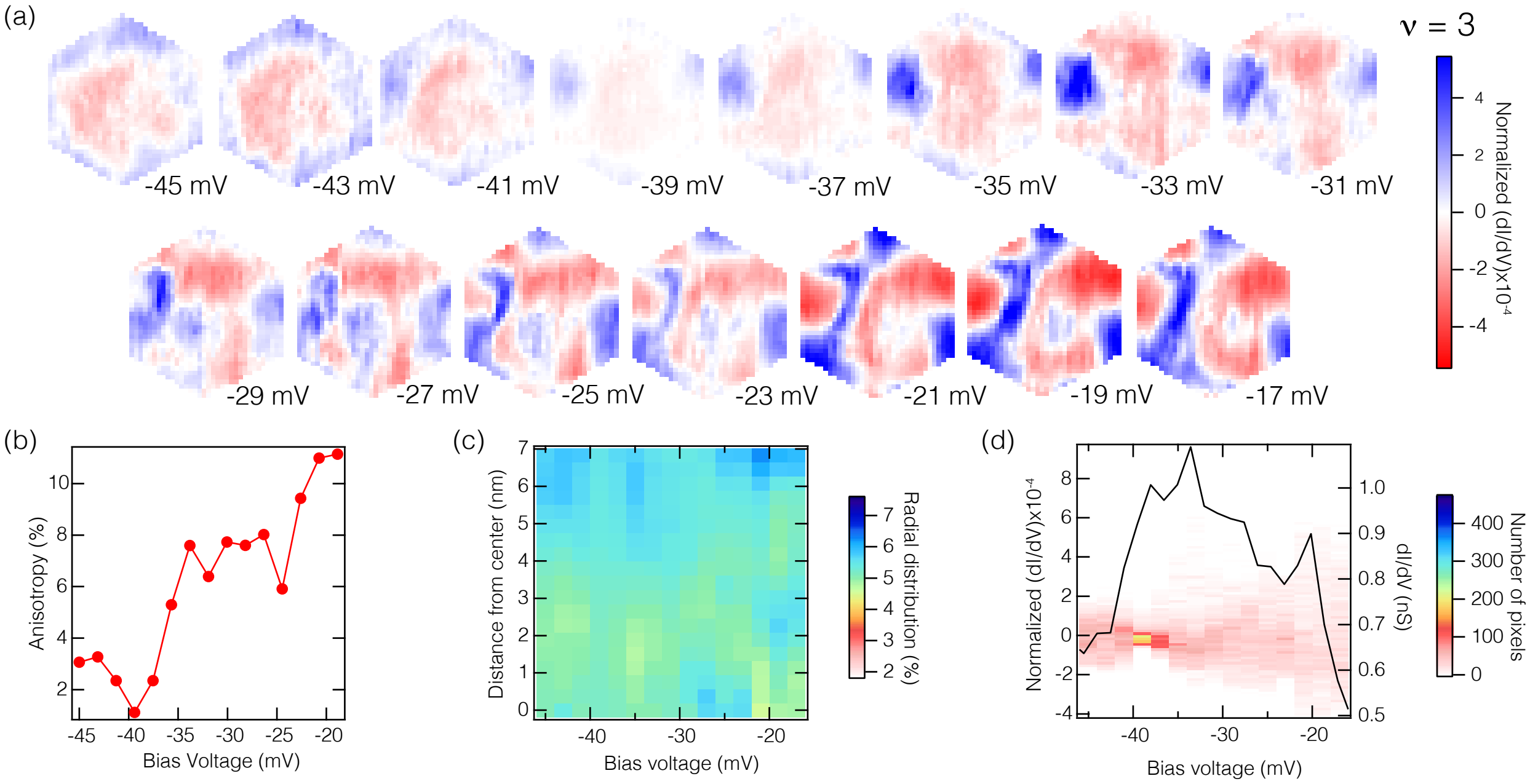

FIG. 13. (a) LDOS maps at different energies for ν = 3. (b)Anisotropy extracted from (a) as a function of bias voltage. (c) Radial distribution extracted from (a) as a function of bias voltage. (d) Color plot plotted against left axis: histogram of (a), black curve plotted against right axis: STS spectroscopy on the center of AA site.

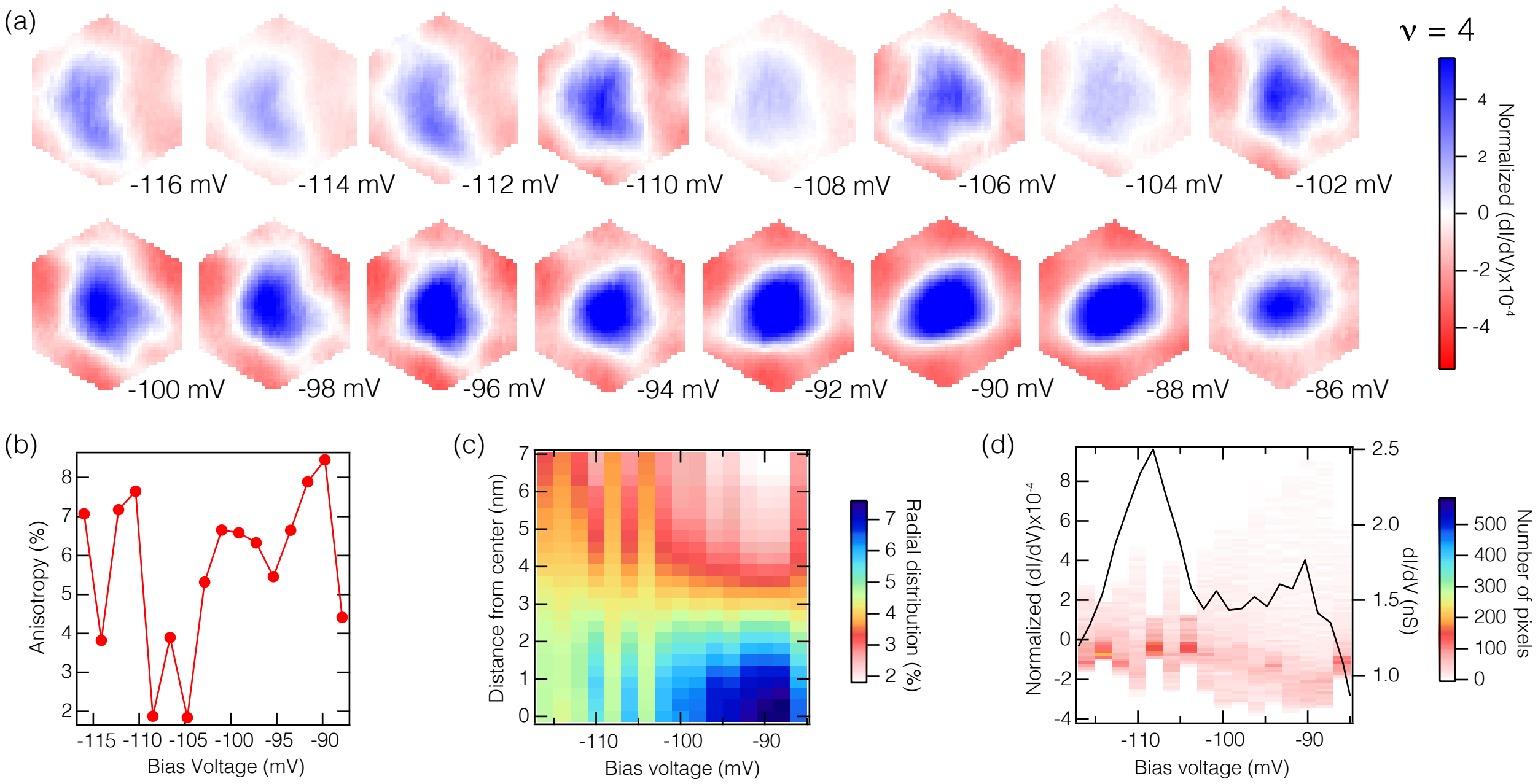

FIG. 14. (a) LDOS maps at different energies for ν = 4. (b)Anisotropy extracted from (a) as a function of bias voltage. (c) Radial distribution extracted from (a) as a function of bias voltage. (d) Color plot plotted against left axis: histogram of (a), black curve plotted against right axis: STS spectroscopy on the center of AA site.

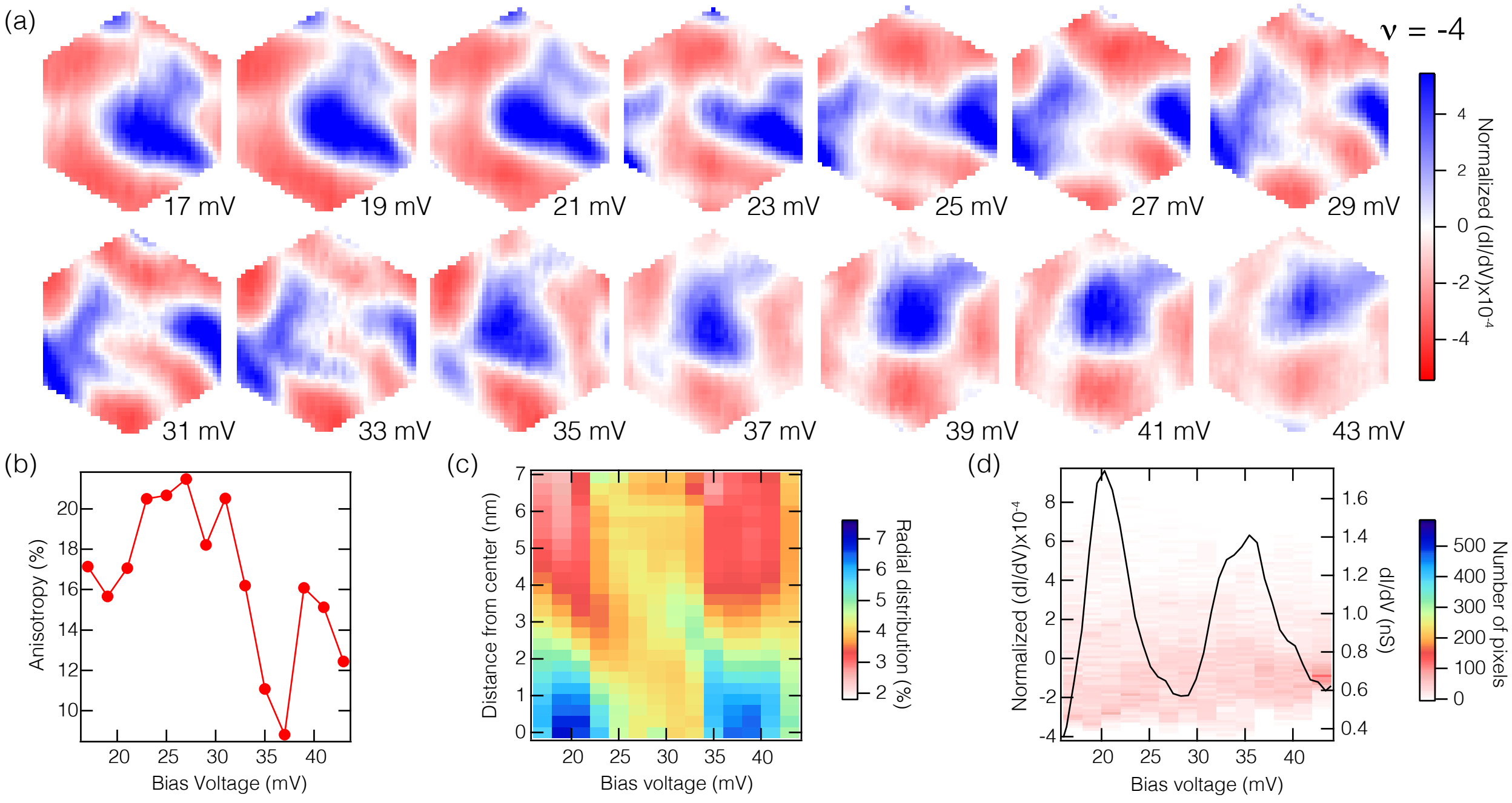

FIG. 15. (a) LDOS maps at different energies for ν = -4. ((b)Anisotropy extracted from (a) as a function of bias voltage. (c) Radial distribution extracted from (a) as a function of bias voltage. (d) Color plot plotted against left axis: histogram of (a), black curve plotted against right axis: STS spectroscopy on the center of AA site.